\documentclass[journal]{vgtc}                

\ifpdf
  \pdfoutput=1\relax                   
  \usepackage{graphicx}                
  \DeclareGraphicsExtensions{.pdf,.png,.jpg,.jpeg} 
\else
  \usepackage{graphicx}                
  \DeclareGraphicsExtensions{.eps}     
\fi%

\graphicspath{{figures/}{pictures/}{images/}{./}} 

\usepackage{microtype}                 
\PassOptionsToPackage{warn}{textcomp}  
\usepackage{textcomp}                  
\usepackage{mathptmx}                  
\usepackage{times}                     
\usepackage{cite}                      
\usepackage{tabu}                      
\usepackage{booktabs}                  

\usepackage{microtype}

\usepackage{ulem}
\makeatletter
\def\ifEmpty#1{\def\@temp{#1}\ifx\@temp\@empty}
\makeatother

\usepackage{xspace}

\usepackage{amsmath,amsfonts,amssymb}
\usepackage{url} 

\newcommand{\FG}[1]{Figure~\ref{#1}}

\newcommand{\shah}{{\textstyle \amalg{\kern-4.pt\amalg}}}

\usepackage{todonotes}
\usepackage{fixltx2e}
\usepackage{dblfloatfix}

\onlineid{1104}

\vgtccategory{Research}
\vgtcpapertype{Algorithm/Technique}

\title{Deadeye Visualization Revisited: Investigation of Preattentiveness and Applicability in Virtual Environments}

\author{Andrey Krekhov, \textit{Student Member, IEEE}, Sebastian Cmentowski, \textit{Student Member, IEEE},\\ Andre Waschk, \textit{Student Member, IEEE}, and Jens Kr\"uger, \textit{Member, IEEE}}
\authorfooter{
\item
 Andrey Krekhov, Sebastian Cmentowski, Andre Waschk and Jens Kr\"uger are with Center of Visual Data Analysis and Computer Graphics (COVIDAG), University of Duisburg-Essen. E-mail: \{andrey.krekhov, sebastian.cmentowski, andre.waschk, jens.krueger\}@uni-due.de.
\item
 Jens Kr\"uger is also with SCI Institute, University of Utah. E-mail: jens@sci.utah.edu.
}

\shortauthortitle{Krekhov \MakeLowercase{\textit{et al.}}: Deadeye Visualization Revisited: Investigation of Preattentiveness and Applicability in Virtual Environments}

\abstract{

Visualizations rely on highlighting to attract and guide our attention. To make an object of interest stand out independently from a number of distractors, the underlying visual cue, e.g., color, has to be preattentive. In our prior work, we introduced Deadeye as an instantly recognizable highlighting technique that works by rendering the target object for one eye only. In contrast to prior approaches, Deadeye excels by not modifying any visual properties of the target. However, in the case of 2D visualizations, the method requires an additional setup to allow dichoptic presentation, which is a considerable drawback. As a follow-up to requests from the community, this paper explores Deadeye as a highlighting technique for 3D visualizations, because such stereoscopic scenarios support dichoptic presentation out of the box. Deadeye suppresses binocular disparities for the target object, so we cannot assume the applicability of our technique as a given fact. With this motivation, the paper presents quantitative evaluations of Deadeye in VR, including configurations with multiple heterogeneous distractors as an important robustness challenge. After confirming the preserved preattentiveness (all average accuracies above 90 \%) under such real-world conditions, we explore VR volume rendering as an example application scenario for Deadeye. We depict a possible workflow for integrating our technique, conduct an exploratory survey to demonstrate benefits and limitations, and finally provide related design implications.

} 


\teaser{
  \centering
  \includegraphics[width=\linewidth]{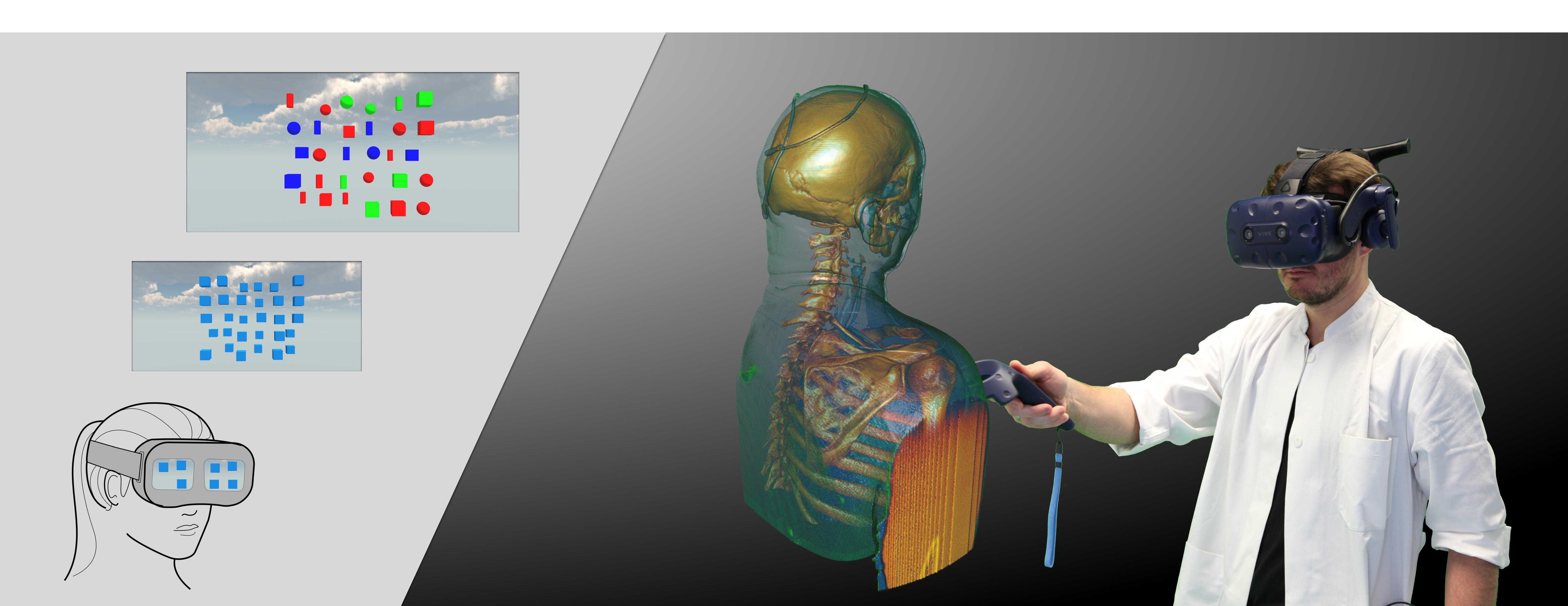}
  \caption{Left: Our experiments in VR with homogeneous and heterogeneous distractors, as we investigate the preattentiveness and robustness of Deadeye in such scenarios. Right: We demonstrate and evaluate volume rendering in VR as a possible real-world application scenario for our technique.}
	\label{fig:teaser}
}

\vgtcinsertpkg

\begin{document}


\firstsection{Introduction}

\maketitle


\begin{figure*}[t!]
\centering
\includegraphics[width=2\columnwidth]{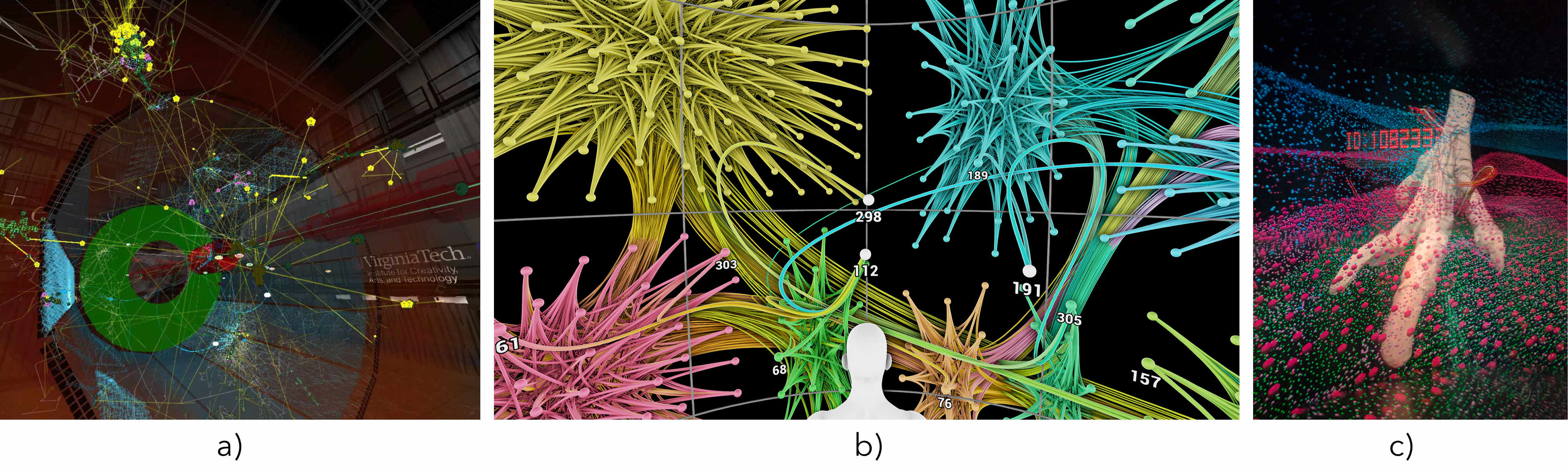}
\caption{A selection of VR visualizations that can benefit from Deadeye highlighting. (a) Educational visualizations of particle physics~\protect\cite{duer2018belle2vr}: Deadeye can be used to capture and guide the attention of the students. (b) Immersive graph visualizations~\protect\cite{kwon16imsv}: Utilizing Deadeye during user interaction to highlight the selected vertices and edges. (c) Dinosaur track formation~\protect\cite{novotny2019developing}: Emphasizing 3D pathlines of interest in unsteady flow visualizations.}
\label{fig:examples}
\end{figure*}

Highlighting objects of interest in scientific visualizations allows our attention to be attracted and guided. Researchers have uncovered a number of effective ways to make an object pop out, be it color, motion, or flickering. Such visual cues can be recognized by our visual system preattentively, i.e., within a split second, no matter how complex the overall visualization. Hence, advances in the exploration of preattentive features can provide substantial benefits to visualization researchers and practitioners.

Our prior work introduced \textit{Deadeye}~\cite{krekhov2019deadeye}, a preattentive visualization technique based on dichoptic presentation. Deadeye attracts and guides attention by rendering the target object for one eye only, which works preattentively due to the induced binocular rivalry. One particular benefit to visualizations is that Deadeye does not modify any visual properties of the target object in contrast to other preattentive cues that have to alter the target by recoloring, reshaping, or displacing (3D depth cue). Clearly, such alterations are undesirable, as they might lead to data misinterpretation. Also, by using Deadeye, we do not have to reserve visual dimensions, such as color for highlighting, and can use these variables for, e.g., encoding more data dimensions. Our studies also confirmed that the display of inconsistent stimuli for each eye does not lead to headache or any other physical strain, which is an important consideration for real-world applicability.

To render the target for one eye only, Deadeye relies on a stereoscopic setup such as a 3D TV with stereo glasses. The only difference between the left and right image is the presence or absence of the target. Hence, one might consider such a setup solely for highlighting purposes as an overkill, and several members of our community proposed applying Deadeye in a truly stereoscopic environment. In such a case, our technique would not require any additional hardware and would perform out of the box. Furthermore, due to the availability nowadays of consumer VR equipment, visualization research increasingly employs virtual environments as a target scenario~\cite{kratz2006gpu,shen2008medvis,laha2012effects,hanel2016visual,scholl,milan2018extending}, and having more complex visualizations in VR triggers the need for suitable highlighting approaches.

Given that community-driven motivation, we revisit Deadeye under stereoscopic conditions. In contrast to more established preattentive cues such as color or flickering, it is rather difficult to predict the behavior of Deadeye in such a setup due to its monocular nature. By removing the object for one eye, we lose the binocular disparity information for the target, which is a crucial part of depth estimation. However, the disparity-based mechanism is not the single point of failure, as our vision system relies on numerous mechanics, such as occlusion geometry~\cite{tsirlin2012vinci}, to estimate the depth of an object, which suggests a successful application of Deadeye in VR scenarios.

\subsection{Motivation for Extending Deadeye to VR}

There are manifold reasons for visualization in VR, such as a better understanding of spatial relationships~\cite{schuchardt2007benefits}, or the increased presence and an improved cognitive map due to natural locomotion (i.e., walking)~\cite{ruddle2011walking,Krekhov:2018:GVRA,ruddle2009benefits}. Nevertheless, these advantages cannot prevent us from possibly getting lost in the amount of visualized data, and, thus, we continue to need robust and intuitive highlighting techniques. And while certain preattentive cues, such as color, are not affected by the transition to VR, temporal approaches, such as flickering, often interfere with aliasing caused by constant micromovements in VR. Hence, we postulate that establishing Deadeye as a validated highlighting approach in VR without occupying any additional visual dimension offers significant benefits for the visualization community, as outlined in \FG{fig:examples}.

One possible application scenario of Deadeye is the educational context. Instructors in virtual reality classes need a subtle, yet efficient way to draw the students' attention to a certain element or an area, such as a particle shower caused by a near-light-speed collision in the Belle II experiment~\cite{duer2018belle2vr}. Similarly, we can utilize Deadeye to highlight particles or pathlines in flow field visualizations~\cite{novotny2019developing}, as our method is not limited to a single target. The only requirement is that the highlighted property, target, or region need to be binary in their nature, as Deadeye either shows or removes it for one eye.

Such highlighting is also beneficial for visual storytelling, be it for experts or the general audience. One example is the biological exploration of cells in VR to better understand the cellular processes~\cite{johnston2018journey}. Again, Deadeye can be used to emphasize various points of interest, such as the nucleus or endosomes, and help the audience to follow their movement during the visualization.

Another candidate for our method would be visualizations that involve cluttered, complex graphs. One important reason for such application scenarios is that VR usually allows to perceive and analyze larger graphs compared to non-stereo setups~\cite{ware1996evaluating}. Naturally, aspects like layout and interaction~\cite{kwon2015spherical,kwon2016study,kwon16imsv} are the dominant factors that impact the success of such visualizations. However, as color is often over-represented in these scenarios, we postulate that Deadeye is a valuable alternative to, e.g., highlight the vertices or edges being currently selected by the user. To summarize, we assume that the unique selling point of Deadeye, i.e., the fact that it does not require a dedicated visual dimension, such as color or motion, is of enough value for a number of VR visualizations, which justifies our follow-on research.

\begin{figure*}[t!]
\centering
\includegraphics[width=2\columnwidth]{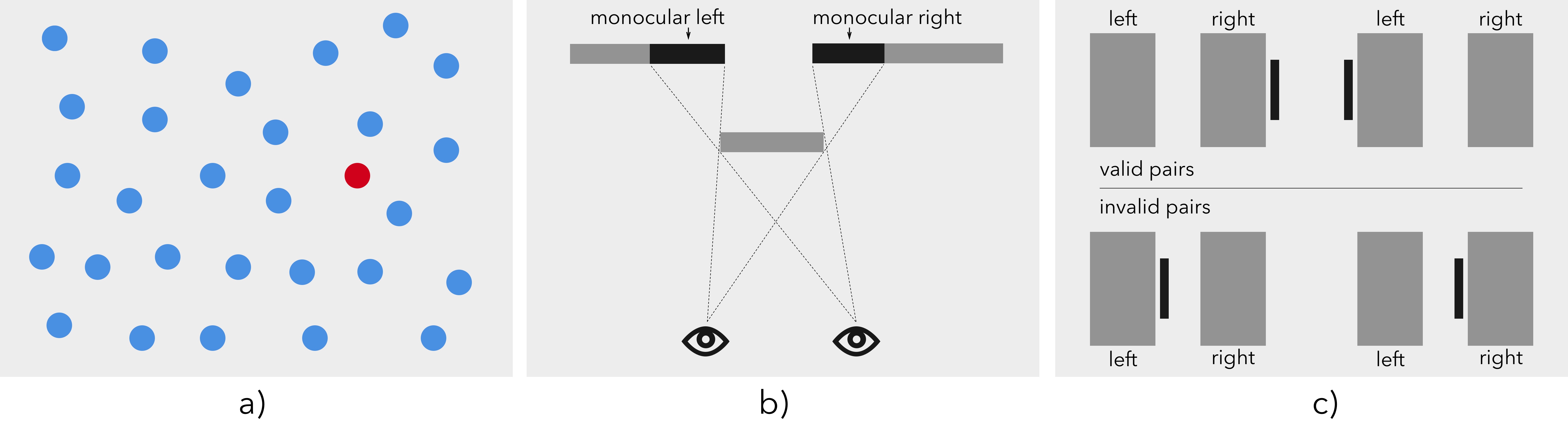}
\caption{(a) Color as a cue: The red circle can be recognized preattentively independent of the number of blue distractors. (b) Da Vinci stereopsis: The near surface results in different occlusions for both eyes, which leads to monocular regions in the far plane. (c) Valid and invalid setups of unpaired image points in da Vinci stereopsis according to Nakayama and Shimojo\protect\cite{NAKAYAMA19901811}. Images redrawn from \protect\cite{ASSEE20072585} and \protect\cite{krekhov2019deadeye}.}
\label{fig:relwork}
\end{figure*}

\subsection{Key Advances of the Follow-On Research}

In this section, we briefly outline the key novelties to facilitate comprehension. Especially readers that are familiar with the basic idea of Deadeye~\cite{krekhov2019deadeye} will find four key advances in this paper:

\textbf{Preattentiveness in VR.} As a first step, we explore whether Deadeye is still preattentive in a VR setup, despite being considered a subtle cue~\cite{zou2017binocularity}. Apart from the listed benefits for VR visualizations, having Deadeye in VR would also remove its prior disadvantage of requiring extra stereo hardware for highlighting.

\textbf{Heterogeneous Distractors.} The original studies on Deadeye were executed with one type of objects, i.e., colored circles. And while this is a valid and most commonly used approach for detecting preattentive cues, it is not guaranteed that these cues perform equally well in a heterogeneous setup with varying distractors. As visualizations are rarely limited to homogeneous objects, we need to assure that the method maintains its robustness under such circumstances. Therefore, we extend the initial research scope by studying the performance of Deadeye under different combinations of heterogeneous distractors: color, shape, and 3D depth.

\textbf{Deadeye and depth perception.} As Deadeye removes the object for one eye, our visual system cannot rely on binocular disparity for depth estimation of the target object anymore. We examine whether our visual system is still able to extrapolate the depth based on additional cues, such as occlusion geometry. This is an important detail, as otherwise, Deadeye might limit our understanding of spatial relationships in VR.

\textbf{Evaluation of real-world applicability.} We demonstrate how to integrate Deadeye into complex scientific visualizations in VR using the example of volume rendering (cf. \FG{fig:teaser}). In particular, we present a workflow for an intuitive application of such a highlighting feature and conduct an exploratory survey with visualization practitioners. Based on the survey outcomes, we discuss the benefits and drawbacks of Deadeye in such scenarios and generate a set of design implications for future research and applications.

\section{Preattentive Cues}

Before exploring the behavior of Deadeye~\cite{krekhov2019deadeye} in a virtual, stereoscopic environment, we briefly introduce the essence of preattentive features and provide a motivation for such research. We will not go into detail regarding the basics of our visual system~\cite{NOTON1971929,itti2001computational,yarbus1967eye} in order to maintain focus on the visualization-related aspects. We recommend the state-of-the-art summary by Healey and Enns~\cite{Healey:2012:AVM:2225054.2225226} for a broad overview of preattentive research done in past decades.

In short, our visual system is able to detect certain outstanding features such as color in a glance, i.e., before our eyes initiate a saccadic movement. Consider the example in \FG{fig:relwork}: looking at such an image for a split second would suffice to tell whether or not there was a red circle among blue ones. Since a saccade usually needs about 200-250 ms~\cite{Healey:2012:AVM:2225054.2225226} to initiate, researchers utilize that threshold to determine if a cue is preattentive. 

Apart from color, prior work has already discovered a large set of other preattentive features, including size, density, lighting direction, flicker, or orientation. For a more detailed overview, we refer to the work by Wolfe and Horowitz~\cite{wolfe2017five} and Healey and Enns~\cite{Healey:2012:AVM:2225054.2225226}. At this point, we also emphasize that preattentive cues differ in their underlying nature, and, thus, the performance or detectability usually depends on various factors, such as the type and heterogeneity of distractors, viewing angle, or lighting condition.

A straightforward application for preattentive cues in visualization is to draw and guide attention~\cite{ware2012information,Hall:2016:FEI,Borji:2013:SAV} due to the nearly instant recognition time. Here, an even more important advantage is that preattentive features perform equally well, no matter how many other objects---also called distractors---are present. In other words, we can confirm a red circle from the previous example in less than 250 ms, even if hundreds of blue elements are on the screen.

Our community has made extensive use of preattentive cues for different approaches, such as dynamic narrative visualizations in the case of \textit{Attractive Flicker}~\cite{Waldner:2014:AFG} or document representations as done with \textit{Popout Prism}~\cite{Suh:2002:PPA:503376.503422}. Other examples include the shading-effects-based \textit{Stylized Focus}~\cite{Cole:2006:DGM:2383894.2383942} and the utilization of 3D depth in graph visualizations~\cite{Alper:2011:SHG:2068462.2068634}. Readers seeking further application examples and higher level design guidelines for such visual features might also be interested in the work by Huber and Healey~\cite{huber2005visualizing}. In addition, studies by Gutwin et al.~\cite{Gutwin:2017:PPI:3025453.3025984} provide insights into performance differences of preattentive cues in the peripheral sight area. Note that Deadeye also suffers from decreasing performance in areas far from the focus point~\cite{krekhov2019deadeye}, and, if this aspect is of high importance (e.g., multimonitor setups), other cues such as motion or flickering should be favored instead.

Although we can spot a single cue instantly, searching for a combination of multiple features often results in a serial and no longer parallel process~\cite{treisman1980feature, treisman1988feature, treisman1986illusory,wolfe1989guided,townsend1990serial,mcleod1988visual}. A detailed discussion of that so-called conjunction search is outside the scope of this paper, yet it is important to know about this limitation of preattentive cues when designing complex visualizations. This rule has a few exceptions, i.e., conjunction search setups where parallel processing can be achieved. Prominent examples are the works by M\"uller and Muhlenen~\cite{muller1999visual} (form and motion) and Nakayama Silverman~\cite{nakayama1986serial} (3D depth and color/motion). That latter report was the main motivation for studying conjunction search abilities of Deadeye, since 3D depth and Deadeye both rely on binocular disparities. However, the experiments did not support the hypothesis that Deadeye is suitable for parallel processing when combined with color. Hence, we are not further examining conjunction search in our VR setup, as we do not see any evidence for a change in behavior in this regard.

To fully understand the underlying mechanisms behind conjunction search and preattentive processing in general, we recommend starting with the Integration Theory by Treisman~\cite{treisman1980feature}. Other fundamental research regarding preattentive models includes the Texton Theory by Julesz et al.~\cite{julesz1981textons} and the Boolean map theory by Huang and Pashler~\cite{huang2007boolean}. As we will not go into detailed explanations of these models, the state-of-the-art paper by Healey and Enns~\cite{Healey:2012:AVM:2225054.2225226} might be considered as a starting point for obtaining an overview.

\begin{figure*}[t!]
\centering
\includegraphics[width=2\columnwidth]{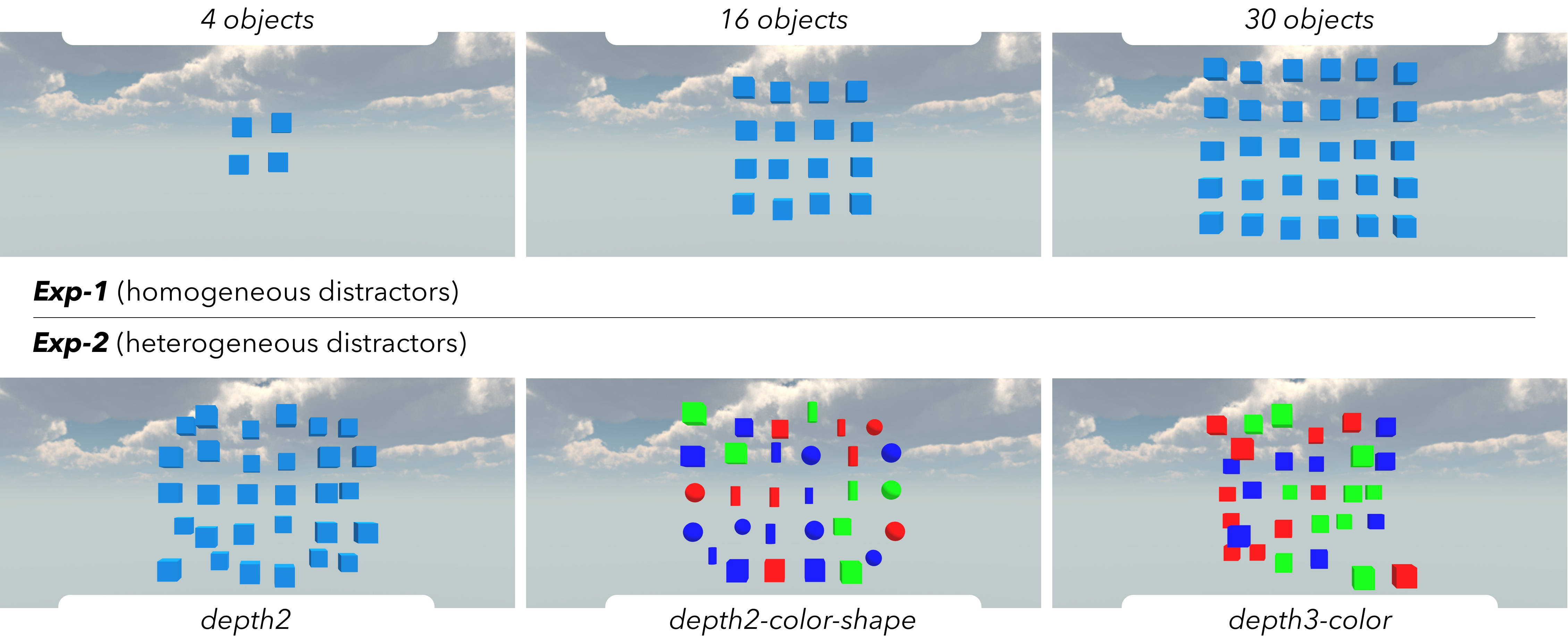}
\caption{ Top: \textit{Exp-1} with an increasing number of homogeneous distractors on the same depth plane with a slight positional jittering applied to each cube. Bottom: \textit{Exp-2} with heterogeneous distractors, from left to right: \textit{depth2} (two depth planes), \textit{depth2-color-shape} (two depth planes, different color and shape), and \textit{depth3-color} (three depth planes, different color).}
\label{fig:preattentive}
\end{figure*}

\section{Underlying Mechanics of Deadeye}

The concept behind the original Deadeye visualization technique is to highlight an object by removing it for one eye only. We refer to such a principle when each eye is exposed to a different stimulus as \textit{dichoptic presentation}. In general, that difference in stimuli leads to binocular rivalry~\cite{logothetis1996rivalling, blake1989neural, friedenberg2012visual, alais2005binocular, Paffen2011}, i.e., our vision system enters a context-switching mode that allows us to perceive both monocular images alternately instead of experiencing a superimposition.

Whether or not binocular rivalry can be perceived in a preattentive manner has been discussed in a number of prior works. Wolfe and Franzel~\cite{wolfe1988binocularity} assumed that such a cue is not preattentive in general, with the exception of the so-called luster effect where the target object is dimmer than the background for one eye and brighter for the other eye. Such luminance variations and luminance disparities in general were subjects of further extensive research~\cite{de1974binocular,teller1967brightnesses,anstis1998nonlinear,formankiewicz2009psychophysics}. Zou et al.~\cite{zou2017binocularity} reported that although dichoptic presentation can be preattentive, it is usually too weak and overridden by more pronounced features such as orientation. Consequently, dichoptic presentation has only rarely been employed in visualization or for highlighting purposes in general~\cite{Zhang:2012:BSE,Zhang:2014:SBE}. 

On the other hand, research by Paffen et al.~\cite{paffen2012interocular} and especially the work by Zhaoping~\cite{zhaoping2008attention} has provided further evidence that binocular rivalry should be reconsidered as a preattentive cue. In particular, Zhaoping confirmed that ocular discontinuities can be used for drawing attention by comparing ocular singletons to orientation singletons and evaluating the role of the primary visual cortex during the construction of related bottom-up saliency maps. The original studies on Deadeye also align with these findings and confirm the achievable preattentiveness.

A well-known case of dichoptic presentation is the binocular disparity generated by the horizontal offset of our eyes. That information is processed by our visual system to gather depth information and forms the basis for stereo vision~\cite{julesz1960binocular,julesz1971foundations,Caziot:2015:SOM,marr1976cooperative,marr1979computational}. On a side note, a number of related stereo cues such as lighting direction and three-dimensionality have also been proven to be preattentive~\cite{enns1990influence, enns1990sensitivity,o1997preattentive}.

In contrast to the disparity-based depth perception, ~\textit{da Vinci stereopsis}~\cite{NAKAYAMA19901811} provides depth information based on monocular vision. This phenomenon is common in daily life, i.e., when an object is partially occluded by another, our vision system perceives monocularly occluded regions as depicted in \FG{fig:relwork}. Although receiving little attention by the general public, monocular occlusion plays an important role in our depth perception process~\cite{gillam1999quantitative,liu1997binocular,shimojo1990real,shimojo1994interocularly,harris2009role}. Consequently, prior research~\cite{marr1979computational,cao2005laminar,hayashi2004integrative,zitnick2000cooperative,watanabe1999stereo} established several models that show how such unpaired image points contribute to depth estimation.

Regarding Deadeye, one important aspect of da Vinci stereopsis is the classification of monocular areas in valid and invalid combinations as shown in \FG{fig:relwork}. Clearly, only valid cases appear in nature. As shown by Shimojo and Nakayama~\cite{shimojo1990real}, such valid regions are not subject to binocular rivalry and usually appear as part of more distant surfaces, whereas invalid regions appear more ambiguous in depth estimation. In the case of Deadeye, the exposed image pair falls into the invalid category. Hence, at first glance, depth estimation for the highlighted object might be error prone. However, more recent research~\cite{HAKKINEN19963815,ASSEE20072585,gillam2003monocular} has revealed that da Vinci stereopsis works in a rather stimulus-dependent way, i.e., the quantitative depth computation approach depends on the given occlusion configuration. Tsirlin et al.~\cite{tsirlin2012vinci} concluded that occlusion geometry is most likely the main source for depth extraction in such cases. These recent findings allow us to assume that Deadeye-enhanced objects would also perform well regarding depth estimation in VR setups, especially when we consider that such a stereoscopic environment allows us to observe the object in question from multiple perspectives to achieve a more sophisticated depth impression~\cite{shimojo1988occlusion}.


\section{Preattentiveness of Deadeye in VR}

We conducted an evaluation to understand whether and how the preattentiveness of Deadeye behaves in a virtual environment. In a first step, we recreated an experiment similar to the original Deadeyes tudy~\cite{krekhov2019deadeye} to generate a set of comparable data. The design of our experiment follows the traditional approach for preattentive cues: series of images---or scenes in our case---are displayed for a short amount of time (100-250 ms), and participants have to decide for each image whether a highlighted object is present or not. If a cue is preattentive, the high success rate (typically $>80$\%) is maintained independently from the number of overall objects on the screen. Hence, preattentive experiments are performed with a varying number of distractors to verify the performance stability.

Accordingly, we formulated a first hypothesis \textbf{H1}: \textit{Stereoscopic environments do not impact the accuracy of Deadeye in the case of \textbf{homogeneous} distractors.} Note that although we utilized stereoscopic equipment in the original experiments to hide the target object for one eye, the image pairs were otherwise identical, i.e., no disparity-based depth cues were present. In contrast, the experiments to be presented utilize a 3D scene that allows spatial vision and where all objects are three-dimensional, as shown in \FG{fig:preattentive}.

In a second step, we significantly extended the original setup by utilizing objects that vary in protruding properties, such as color or shape, to explore the robustness of our technique. Note that preattentive cues perform differently under heterogeneous conditions, and understanding such behavior is crucial because visualizations seldom consist of only one type of objects. Since prior research~\cite{zou2017binocularity} has demonstrated that dichoptic presentation is a rather weak cue that could be easily overridden by stronger features, we regard an evaluation under heterogeneous conditions as an important milestone for establishing Deadeye as a ``working'' cue for real-world visualizations. Hence, our second hypothesis is that \textit{Deadeye accuracy remains robust in scenarios where distractors and the targets have \textbf{heterogeneous} visual properties} (\textbf{H2}).

\begin{figure*}[!t]
\centering
\includegraphics[width=2\columnwidth]{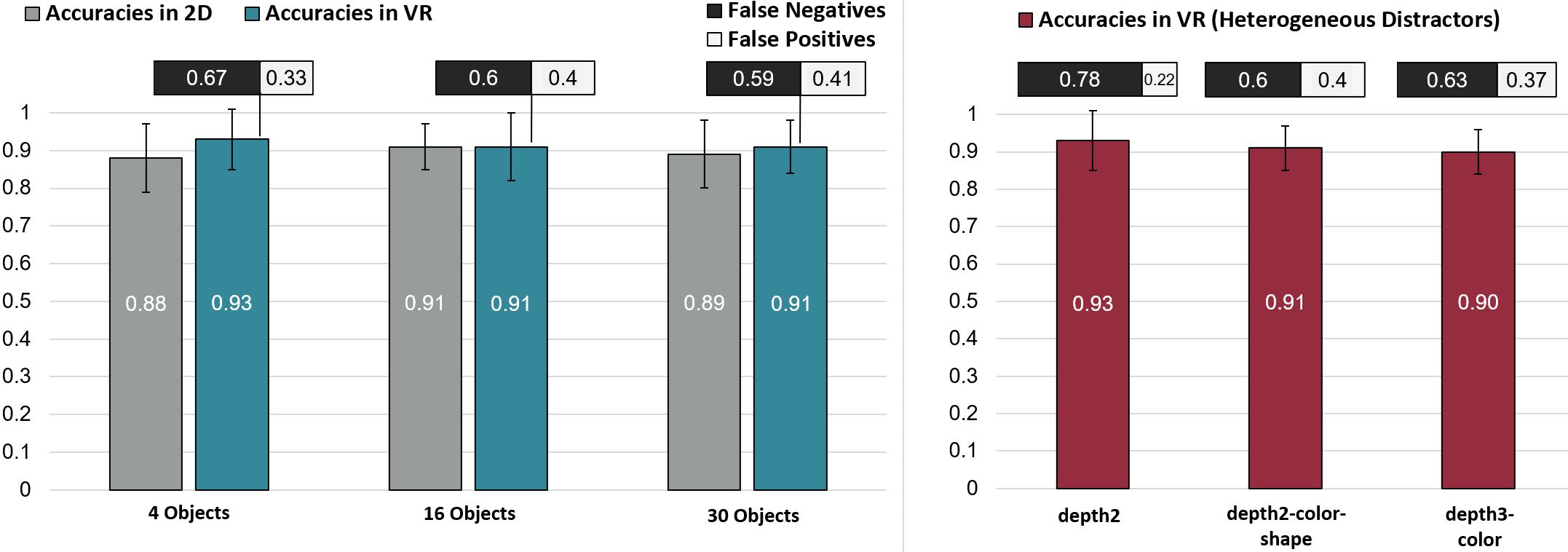}
\caption{Average accuracies for \textit{Exp-1} and \textit{Exp-2} compared to our previous results~\cite{krekhov2019deadeye} for the 2D case. A repeated measures ANOVA shows no significant difference in accuracy between the sets, supporting our hypotheses that the transition to 3D scenes does not impact the performance of Deadeye and that our technique remains robust even in the presence of strong visual cues such as color or shape.}
\label{fig:results}
\end{figure*}


\begin{figure}[!b]
\centering
\includegraphics[width=1.0\columnwidth]{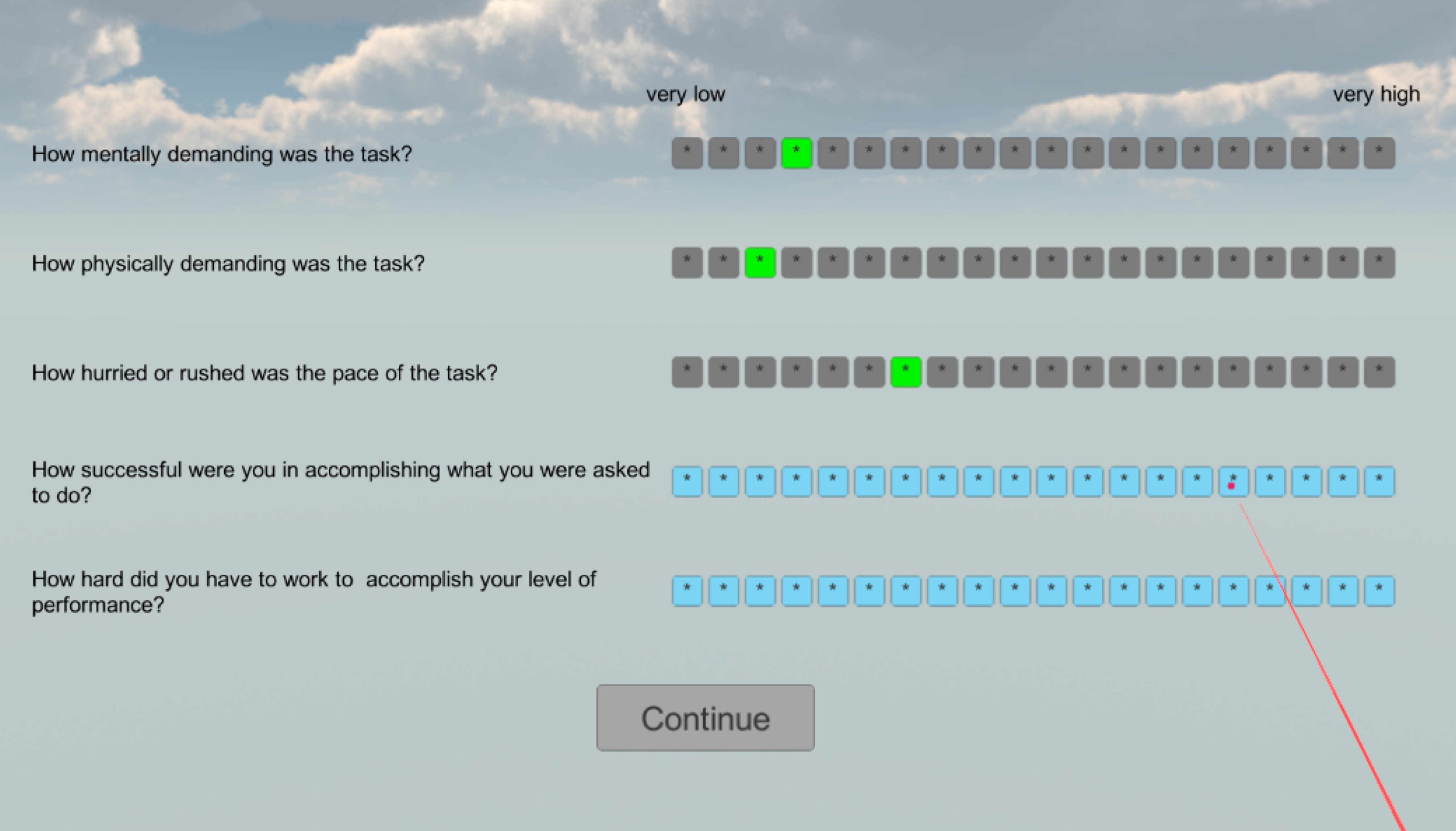}
\caption{Excerpt from our VR implementation of the NASA-TLX survey. All briefings, pauses, and questionnaires were done via the HMD to provide same conditions to all participants.}
\label{fig:questionnaire}
\end{figure}

\subsection{Procedure and Applied Measures}

Our experiments took place in a virtual reality lab. In our recruitment call, we requested normal or corrected to normal visual acuity and no defects of vision as a study prerequisite. Upon entrance, we informed the participants about the overall procedure and administered a questionnaire to obtain general demographic data. 

We then provided the participants an Oculus Go~\cite{oculus} HMD with a per eye resolution of 1280 x 1440 pixel and briefly explained its usage. The remainder of the study took place in the VR environment, i.e., all trials, briefings, pauses, and questionnaires were done via the HMD to guarantee equal conditions (e.g., between-run delays) for all participants.

On-screen text briefed the participants that they would see a series of static 3D scenes consisting of floating objects, and one of the objects might pop out. Each scene was shown for a split second, and, subsequently, participants would vote via \textit{yes} and \textit{no} buttons whether they thought that a highlighted object was present. Before each scene, participants saw a crosshair in the middle of the screen and were asked to maintain focus on it. This detail is important to prevent saccadic eye movements such that the eyes would remain in the focus stage when a scene becomes visible. The briefing also mentioned that there would be a training stage before each round where participants would receive audio feedback that indicated the correctness of the given answer. During the real run, the audio would be replaced by a neutral sound to prevent distractions, i.e., reflecting upon wrong answers and lack of concentration as a result.

The entire study consisted of two parts: \textit{Exp-1} with homogeneous distractors and \textit{Exp-2} focusing on heterogeneous distractors. \textit{Exp-1} consisted of three sets that differed only in the number of displayed objects: 4, 16, and 30. We utilized randomly generated cubes on a 5 x 6 grid
with jittering/offset functions as depicted in \FG{fig:preattentive}. All cubes were aligned on the same depth plane, resulting in a horizontal viewing angle of $14.93^\circ$ from the focus point (vertical: $12.23^\circ$). Each cube had a size of approximately $1.8^\circ$. Each of the three sets included 48 scenes, half of them with a target object at a random position in a randomized order. We exposed each scene for 250 ms to the participants; the previously displayed crosshair lasted for 2500 ms. Furthermore, each set began with a training round with 20 scenes. To summarize \textit{Exp-1}, we replicated our setup from the previous study as precisely as possible to generate statistically comparable data. The only altered condition was the shift from a 2D image with flat objects to a scene with three-dimensional objects with binocular disparity.

\textit{Exp-2} included the three sets \textit{depth2}, \textit{depth2-color-shape}, and \textit{depth3-color} in random order. All three sets consisted of 30 objects and varied only in the type and/or depth distribution of utilized elements. All other conditions were similar to \textit{Exp-1} to allow direct comparisons. For \textit{depth2}, we chose the same objects as in \textit{Exp-1}, but distributed them on two distinct depth planes as shown in \FG{fig:preattentive} and also included a minor depth jittering function. The distance between the depth planes was twice the cube-side length. The maximum partial occlusion of a far-plane object by a closer one was 10 \% of its screen space, i.e., at least 90\% of the far object remained visible. For \textit{depth2-color-shape}, we included different forms of objects and also randomly assigned different colors to see how Deadeye performs in the presence of two rather dominant visual cues. In addition, we kept the depth distribution over two planes. The scenes of \textit{depth3-color} further increased the depth variance by adding a third plane and increasing the maximum possible occlusion of the furthermost object to 20 \%. The objects were all cubes with randomly assigned color.

Between all sets of our study, we displayed a stereoscopic 3D landscape photo for 30 seconds to provide short pauses and reduce task fatigue, similar to the original Deadeye experiments. We also re-implemented the original questionnaire in VR, as shown in \FG{fig:questionnaire}, and administered it after \textit{Exp-1} and after \textit{Exp-2}. The questionnaire is based on the NASA-TLX survey~\cite{hart1988development}, which includes six subscales: mental demand (low/high), physical demand (low/high), temporal demand (low/high), performance (good/poor), effort (low/high), and frustration level (low/high). Each scale ranges from 0 to 100 in increments of 5. Our questionnaire also contains two additional items on a seven-point Likert scale ranging from 0 to 6, with larger numbers indicating a more positive outcome: \textit{\textbf{Clearness}: how well have you perceived the highlighted object?} and \textit{\textbf{Decision-making}: how sure were you that you made the right decisions?}

\begin{figure*}[!t]
\centering
\includegraphics[width=2\columnwidth]{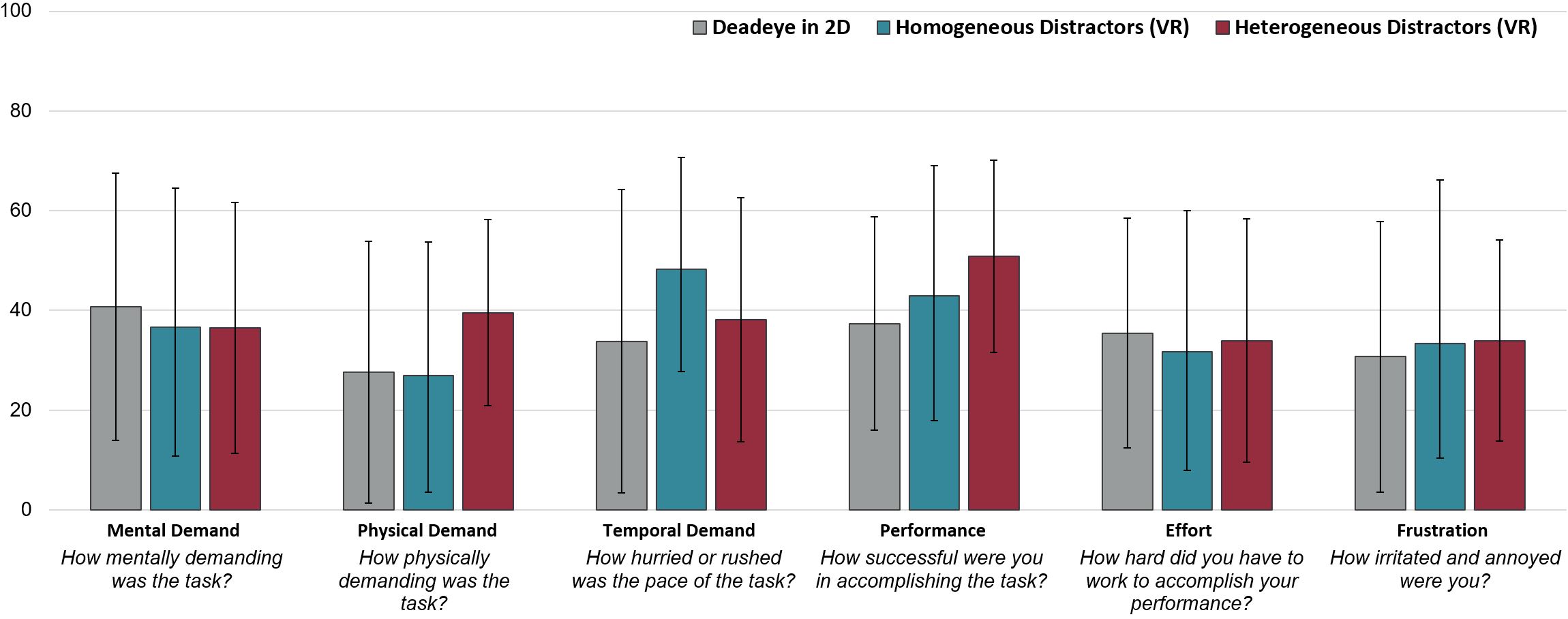}
\caption{ Results of the NASA-TLX survey for \textit{Exp-1} and \textit{Exp-2} compared to the prior values of the 2D case. Lower values are preferable.}
\label{fig:nasa}
\end{figure*}

\begin{figure}[!b]
\centering
\includegraphics[width=0.78\columnwidth]{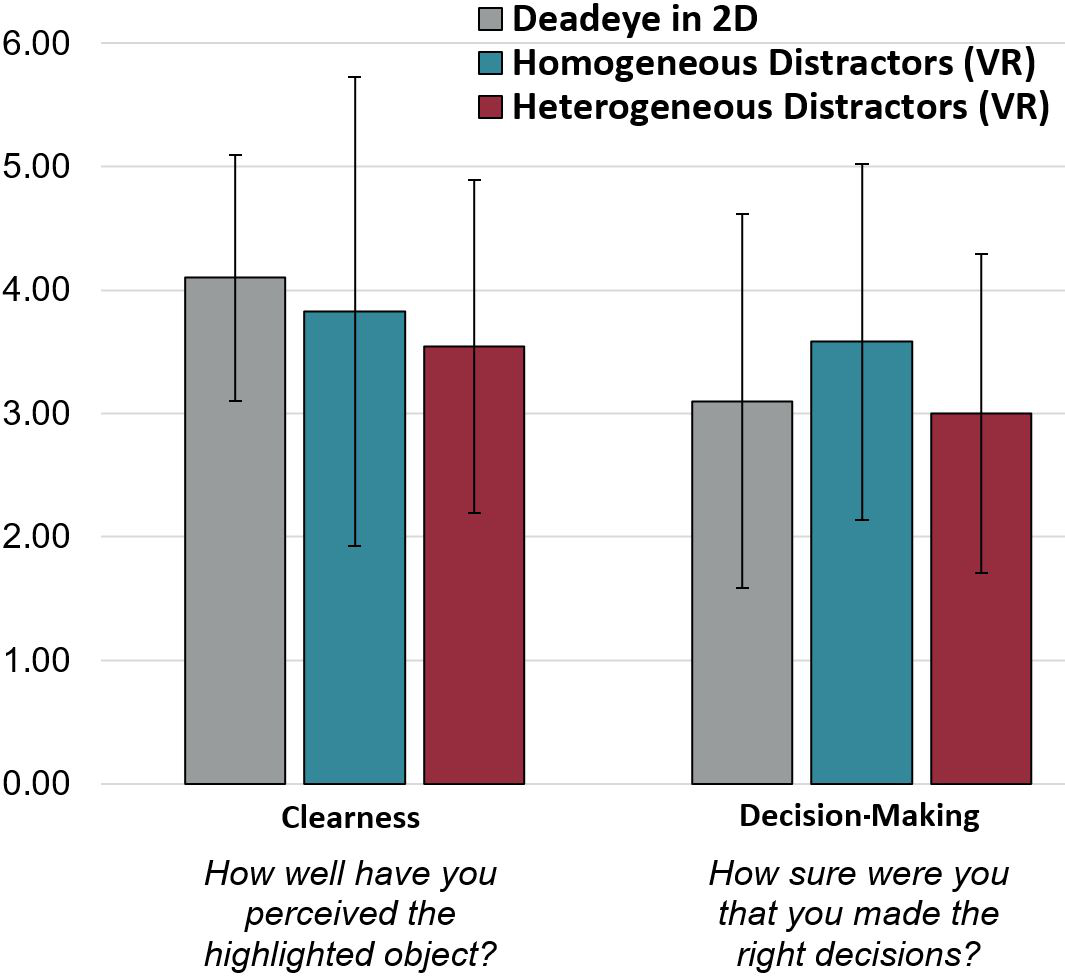}
\caption{Results of our custom questions. Larger numbers indicate a more
positive outcome.}
\label{fig:custom}
\end{figure}

\subsection{Results}

The quantitative study included twenty-four persons (fifteen females, nine males), aged 18 to 43 ($M = 26.71,\,SD = 6.82$). All reported normal or corrected to normal visual acuity and no defects of vision. The presented results are based on the automated logging of our VR application. All variables were normally distributed according to Kolmogorov-Smirnov tests.

The average accuracies, as depicted in \FG{fig:results}, for \textit{Exp-1} (4 objects: $M = 0.93,\,SD = 0.08$; 16 objects: $M = 0.91,\,SD = 0.09$; 30 objects: $M = 0.91,\,SD = 0.07$) and \textit{Exp-2} (\textit{depth2}: $M = 0.93,\,SD = 0.08$; \textit{depth2-color-shape}: $M = 0.91,\,SD = 0.06$; \textit{depth3-color}: $M = 0.90,\,SD = 0.06$) are similar to the values of other preattentive cues. We applied a repeated measures ANOVA with the set type as a within-subject variable to investigate whether the number and kind of distractors influence the accuracy of Deadeye in VR. The result, $F(5,115) = 1.04,\,p = .397$, shows no significant difference. In other words, the performance of the participants was affected by neither increasing the number of cubes nor adding depth layers and modifying the shape and color of displayed objects. Regarding the false negatives to false positives ratio in case of wrong answers as depicted in \FG{fig:results}, it is more likely to miss a target than to report a false alarm, which is also a common behavior of other preattentive cues.

We also performed independent samples t-tests to explore whether the transition to VR had an impact on performance. To enable such comparisons, our sets in \textit{Exp-1} replicated the prior 2D setup~\cite{krekhov2019deadeye} as close as possible. The results for 4 ($t(43) = -1.97, p = .055$), 16 ($t(42.18) = -0.02, p = .988$), and 30 objects ($t(43) = -0.85, p = .402$) show no significant differences in accuracy between 2D and 3D experiments.

To evaluate the subjective perception of Deadeye, we conducted a paired-samples t-test to compare the outcomes of the NASA-TLX questionnaire for \textit{Exp-1} and \textit{Exp-2}, as shown in \FG{fig:nasa}. Physical demand ($t(23) = -2.22, p = .037$) and temporal demand ($t(23) = 2.72, p = .012$) are significantly different. Physical demand was reported to be significantly higher ($M = 39.58,\,SD = 18.65$ vs $M = 26.88,\,SD = 23.35$) in \textit{Exp-2}, whereas temporal demand was significantly higher in \textit{Exp-1} ($M = 48.33,\,SD = 20.62$ vs $M = 38.13,\,SD = 24.44$). Furthermore, we applied independent samples t-tests to compare the NASA-TLX results of \textit{Exp-1} with the prior 2D experiment and found no significant differences between any of the subscales. 


In a similar way, we compared our custom questions regarding clearness (\textit{Exp-1}: $M = 3.83,\,SD = 1.90$; \textit{Exp-2}: $M = 3.54,\,SD = 1.35$) and decision-making (\textit{Exp-1}: $M = 3.58,\,SD = 1.44$; \textit{Exp-2}: $M = 3.00,\,SD = 1.29$) as can be seen in \FG{fig:custom}. We found no significant differences between \textit{Exp-1} and \textit{Exp-2}, or between \textit{Exp-1} and our prior 2D results.


\subsection{Discussion}

Our results support our assumption that a virtual environment does not limit the preattentive nature of Deadeye, because the feature is still recognized with an average accuracy of $\sim$~90\%. In particular, the outcomes of our independent samples t-tests for the sets of \textit{Exp-1} and the respective sets of the prior 2D experiment support our hypothesis \textbf{H1}, i.e., the transition to fully stereoscopic environments does not impact the behavior of Deadeye in the case of \textbf{homogeneous} distractors.

The behavior regarding wrong answers also remains unchanged: about $\sim$~65\% of the errors were false negatives, which is quite common for preattentive cues. In other words, it is easier to overlook a highlighted object rather than it is to mistakenly call out a false alarm in the absence of a highlight.

In our opinion, the most important finding of our experiment is that the performance of Deadeye was not impacted by \textbf{heterogeneous} distractors, because our evaluation did not reveal any significant differences in performance and fully supported our second hypothesis \textbf{H2}. We intentionally picked strong visual attributes for our distractors, such as shape and color, because previous work suggested that the preattentive character of binocular rivalry might be easily overridden~\cite{zou2017binocularity}. However, Deadeye performed surprisingly robustly, which allows us to suggest the technique as a suitable highlighting method for real-world applications. Clearly, an unstable performance under heterogeneous conditions would have significantly limited the potential of Deadeye, because visualizations, in most cases, involve a complex interplay among multiple visual attributes.

Furthermore, we made two rather unexpected observations during \textit{Exp-2}. First, the performance of the \textit{depth3-color} set surpassed our expectations. We suspected that target objects on the most distant depth plane would perform significantly worse due to the small screen size and the partial occlusion. However, as shown in \FG{fig:accuracies}, a detailed analysis of missed target locations did not reveal any error trend toward a specific depth plane. Second, we assumed that thinner objects, i.e., cylinders in the \textit{depth2-color-shape} set, might perform worse due to the influence of aliasing (``shimmering'') caused by slight HMD movement. Yet again, analyzing the missed target locations did not reveal any relation between shape and error rate. The only notable connection between the target position and its likeliness to be overlooked is the angular distance from the focus point, i.e., Deadeye performance decreases for objects in the outer columns. At this point, we will not go into detail regarding this finding, because it was already described and analyzed in the original Deadeye paper~\cite{krekhov2019deadeye}. Furthermore, as the visual quality of current HMDs is very limited in peripheral areas, this aspect is of low importance compared to non-VR, multi-monitor setups. 

Interestingly, a closer inspection of the NASA-TLX questionnaire outcomes does not give us a clear picture regarding the actual exertion induced by Deadeye. Although the average values are positive compared to other visual cues~\cite{Gutwin:2017:PPI:3025453.3025984}, we point our readers at the rather large standard deviations. For each subscale, the answers provided by the participants always contained both extreme values, i.e., 0 and 100. We suppose that such a spread is due to the subtle nature of Deadeye and our inability to put the perceived effect into words. This presumption is also underpinned by our custom questions that reflect an average confidence of participants regarding their made decisions.

As a side note, we attribute the significant differences for physical and mental demand between \textit{Exp-1} and \textit{Exp-2} to possible sequence effects of our study. Temporal demand decreased over time, because participants got used to the fast pace of the trials. In contrast, physical demand increased with the duration of the experiment, because the HMD is still an uncomfortable device when worn over a longer period of time.

\begin{figure}[t!]
\centering
\includegraphics[width=0.98\columnwidth]{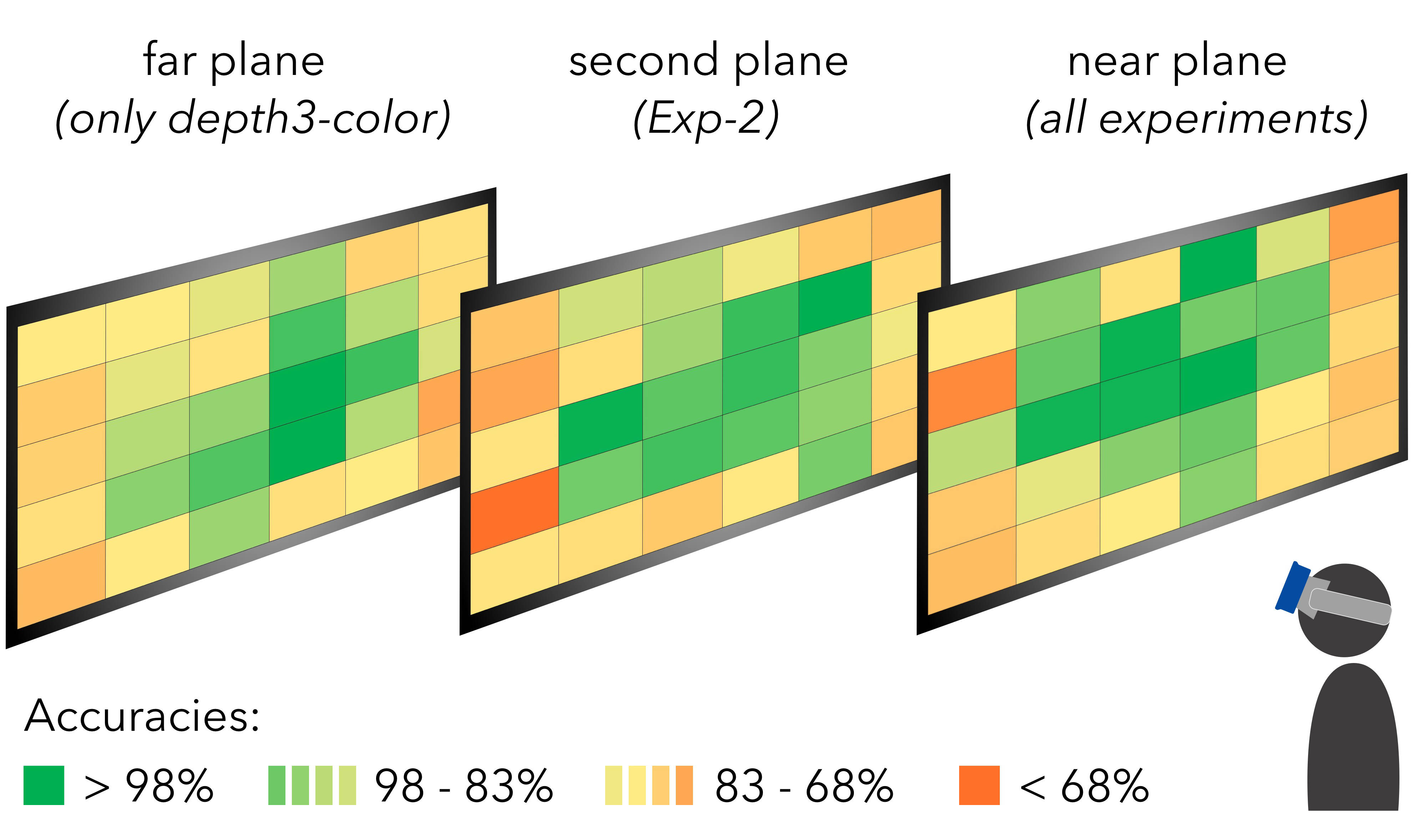}
\caption{Average success rates for the detection of a Deadeye-enhanced
object at each position. The matrices indicate an increase of the error rate with increasing distance from the focus point and no notable difference regarding the target's depth location.}
\label{fig:accuracies}
\end{figure}

\section{Integration into VR Volume Rendering}

In addition to rather fundamental studies on preattentiveness, we now consider volume rendering in VR as a real-world application scenario of Deadeye. We present an integration example and an exploratory study to demonstrate the benefits and limitations of our approach.

\subsection{Implementation}

Given a volume rendering system with VR support (cf. \cite{kratz2006gpu,shen2008medvis}), the technical implementation of Deadeye can be done in the following way. For simplicity, let us assume that Deadeye should be achieved by removing the target for the \textit{right eye}. To remove the footprint of the target volume regions for the right eye, we store a per-voxel mask volume on the GPU along with the volume data. During GPU-based ray-casting (cf. \cite{kruger2003acceleration}), we compare this mask value against a global constant in the shader, which whether the rendering is done for the left or right eye. A voxel is skipped if the mask is set and rendering is done for the right eye. 

The generation of such a mask volume depends on the application domain. For instance, in the case of already segmented~\cite{litjens2017survey} data sets, the per-voxel segment ID of the region of interest can be utilized. Another option is to interactively create and modify the mask volume by ``erasing'' the target for one eye directly in VR, as done in our study.

\subsection{Procedure of the Study}
To gather more information about the behavior of Deadeye in real-world-conditions, we conducted a semistructured exploratory study based on medical volume rendering in VR. Our main goal was to determine possible limitations of our preattentive highlighting method and to generate a preliminary set of design guidelines for the application of Deadeye in such scenarios.

Since we were interested in feedback from people who are already familiar with volume rendering, we decided to recruit our participants manually via direct e-mail requests to limit the discard rate. The experiment took place in our VR lab and was supervised by two researchers. One of them was responsible for briefing and interviewing the participant during the experiment. All participants agreed that we would perform an audio recording of the whole session to facilitate the final evaluation. The second researcher monitored the volume rendering application and executed certain requests from participants, such as resetting the highlighted regions or switching between data sets. We intentionally reduced the direct interactions of the participants with the software to a minimum to focus on the highlighting aspect. 

After assessing the demographic data and conducting a hole-in-the-card-test for eye dominance (Dolman method, e.g., \cite{cheng2004association, porac1976dominant}), we equipped our participants with an HTC Vive Pro~\cite{vive} with a wireless adapter and explained the remainder of the study. For volume rendering, we used a VR-optimized, GPU-based ray caster implementation that guaranteed a minimum refresh rate of 90 frames per second. We used two data sets to simulate common medical situations, a CT scan of a human head (+neck/shoulders), and a CT scan of a human torso; both are subsets from the frozen CT visible human data set~\cite{662875}. In the first part of the study, participants explored both data sets with a medical greyscale transfer function. For the second part, we applied a transfer function that included color to explore its interplay with Deadeye. Each of these four scenarios (2x greyscale, 2x colored; cf. \FG{fig:volume}) consisted of the following tasks:

 \begin{itemize}
  \setlength\itemsep{0.02 em}
  \item \textbf{Detection} - we displayed two different presets (scale, orientation, clip plane); each preset included between one and four Deadeye-enhanced regions. Assignment: \textit{Tell whether and how many parts of the data set pop out. Describe where the highlighted elements lie in depth relative to their surrounding.}
  \item \textbf{Application} - participants were able to apply Deadeye on their own using the trigger button of the controller similar to 3D painting. Assignment: \textit{Pick two regions of interest and highlight them as precisely as possible.}
  \item \textbf{Comparison} - we displayed two presets that utilized alternative popout techniques, color and flickering, to highlight between one and four regions. In the case of colored scenarios, only flickering was applied. Assignment: \textit{Locate the highlighted regions. Describe your perceived benefits and drawbacks compared to Deadeye.}
\end{itemize}

During each assignment, we explicitly asked the participants to provide feedback in a think-aloud manner if possible. We did not impose any time limit for the scenarios and encouraged brief pauses between them. Upon completion of the experiment, we offered the opportunity to provide any additional feedback if desired.

\begin{figure}[t!]
\centering
\includegraphics[width=0.98\columnwidth]{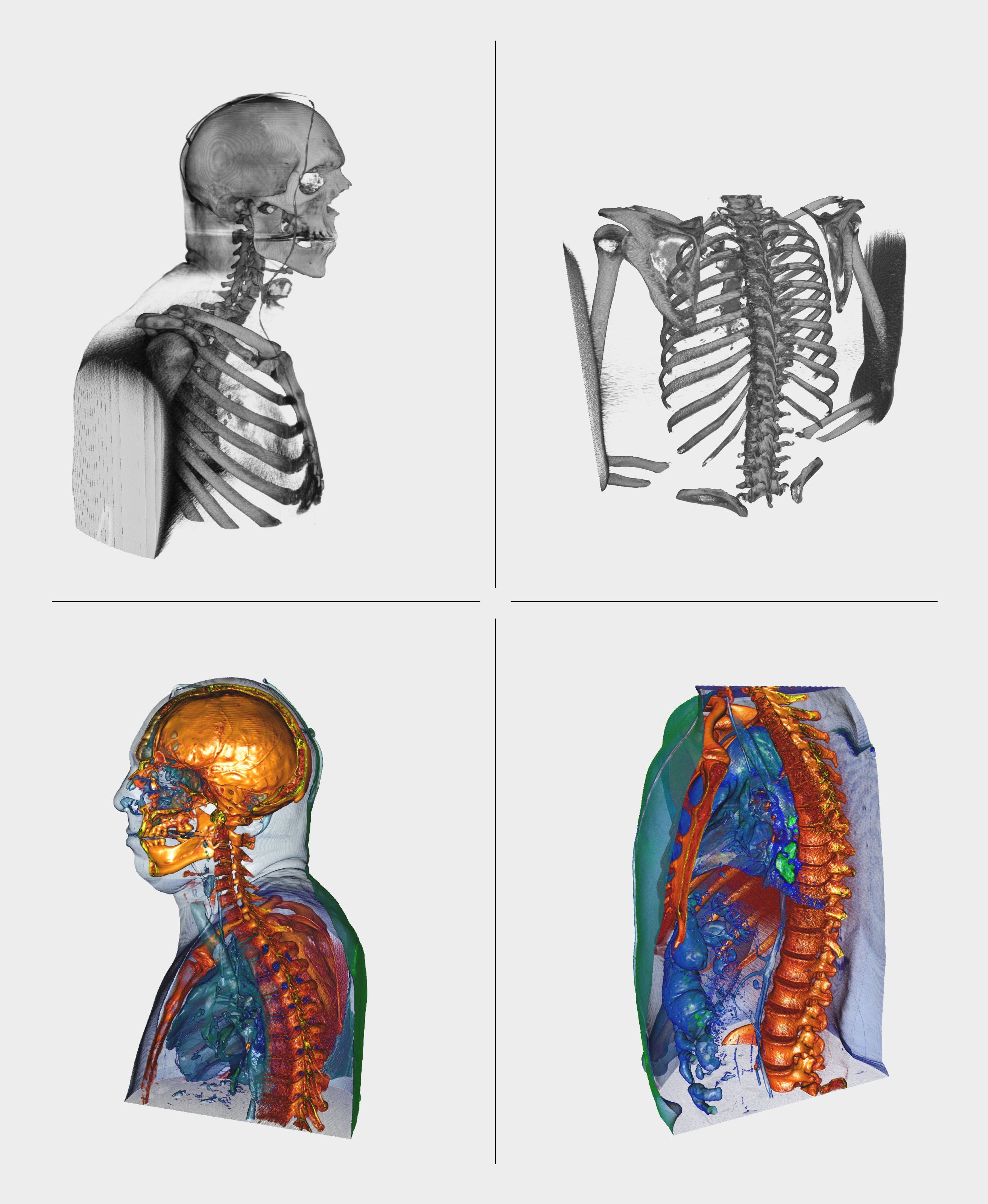}
\caption{Participants went through four different visualization scenarios based on the visible human data set: head including neck and shoulders, and torso, both with greyscale and color transfer functions. Each scenario consisted of three tasks: spotting Deadeye and determining the depth of the target (\textbf{Detection}), interactively applying Deadeye (\textbf{Application}), and comparing Deadeye to flickering and color (in greyscale scenarios only) as alternative highlighting approaches (\textbf{Comparison}).}
\label{fig:volume}
\end{figure}

\subsection{Findings}

The exploratory study included nine participants (three females, six males), aged 29 to 45 ($M = 36.67,\,SD = 5.68$) participated in the exploratory study. All had prior experience with volume rendering, and six of the nine participants had used VR HMDs before. The majority ($N = 7$) had a right dominant eye, and all participants reported normal or corrected to normal visual acuity. In the remainder of this section, we group the obtained results by the respective task to provide a structured exposition.

\textbf{Detection.} In all four scenarios, participants delivered solid performances in spotting and pointing at highlighted regions via a selection ray. Only two participants made a false negative error each. Both participants missed the same small, thin region located rather deep within the data set. Four participants told us that they were \textit{able to spot target regions easier when the data sets were colored (\textbf{P3})}. All participants were able to describe the relative depth of the highlighted objects relative to the surrounding elements. Four participants stated that they had to \textit{move the head a little bit to view the object from different angles to be 100 \% sure (\textbf{P7})}.

We received differing comments regarding the perception of the highlighted object. All but one participant utilized the adjective \textit{transparent} in certain variations, e.g., \textit{pseudo-transparent (\textbf{P1})} or \textit{semitransparent (\textbf{P6})}. All participants stated that they \textit{could see the target as well as the objects behind it (\textbf{P3})}. Seven participants emphasized that they \textit{seemed to be able to decide whether they want to ``see'' the highlighted region or suppress it by focusing on the background objects (\textbf{P4})}. Note that this observation aligns with general binocular rivalry behavior where we perceive both monocular images alternately instead of seeing a superimposition. Two participants were particularly intrigued by that behavior, describing it as \textit{a kind of willpower-dependent rendering where one could mentally alter the perceived visualization similar to ambiguous image puzzles (\textbf{P8})}. Three participants also noticed that one \textit{could modify the transparency level of the highlighted object by rotating the head (\textbf{P9})}. This phenomenon is caused by the rather limited field of view of the HMD, i.e., when we look at an object from a sufficiently large angle, that area is no more present for the eye that is further away. Hence, binocular rivalry becomes disabled, and we perceive a valid monocular image from the nearer eye that either contains the object or not.

At the end of the detection assignment, we also varied the eye for which the object would be removed. However, participants were indecisive, and only three of them expressed a weak tendency toward the object removal for their dominant eye. Other participants stated that they \textit{felt a slight difference, but cannot say which option is better (\textbf{P4})}.

\textbf{Application.} All participants were able to manually apply Deadeye onto self-chosen regions of interest. However, the majority ($N = 8$) perceived the task as rather \textit{challenging especially when moving the controller in the depth dimension (\textbf{P2})}. One participant compared it to a VR painting tool: \textit{It feels the same as 3D painting in VR, which was not very intuitive to me, either (\textbf{P5})}. Seven participants told us that they \textit{intuitively closed one eye from time to time to verify the result and speed up the highlighting because it feels more controlled that way (\textbf{P3})}. No participants attested to any negative feelings about that one-eye trick. As expected, all participants stated that applying Deadeye is easier for colored data sets because one \textit{could just pick a region of a specific color which already has a clear visual separation from the surroundings (\textbf{P1})}.

\textbf{Comparison.} In the case of greyscale scenarios, all participants preferred color over Deadeye for two reasons. First, color as highlight is \textit{more prominent, straightforward, and already known (\textbf{P6})}. Second, as pointed out by six participants, color, in contrast to Deadeye, is not binary. One could \textit{utilize a meaningful color scheme for highlighting different regions and even have gradients and interpolations if needed (\textbf{P9})}. Deadeye instead \textit{is either active or not, without any possibility of providing an additional meaning or more granular differentiation (\textbf{P7})}. However, color is not applicable as a preattentive highlighting method in the case of colored data sets, which is an important limitation compared to Deadeye.

When comparing Deadeye to flickering as a highlighting method, the majority ($N = 8$) preferred Deadeye; one participant remained indecisive. The most prominent reason for that preference was that flickering \textit{is often affected by aliasing and other artifacts (\textbf{P2})} that occur because such a VR experience is never a static scene due to the permanent view changes. Hence, the preattentiveness of temporal effects such as animation or flickering can be rather weakened in such setups, whereas the performance of Deadeye remains stable.

\subsection{Discussion and Design Implications}

The results of both studies support our vision that Deadeye is applicable as a preattentive visualization technique in real-world scenarios. Even in complex visualizations such as volume-rendered CT scans, we can still recognize the highlighted region and identify its depth. Although disparity-based depth perception is limited, our vision system is still capable of utilizing less straightforward cues such as occlusion geometry. Furthermore, if in doubt, we can easily refine our depth estimation of a region by slightly moving our head to generate different view angles.

One interesting aspect of Deadeye is that we can look behind the highlighted target by focusing on objects that lie behind it. This multistable perception---induced by binocular rivalry---can be regarded as a benefit or drawback depending on the use case. On one hand, such conditional perception allows us to gather additional information without any interaction with the underlying system. In particular, we think of scenarios with very limited interaction possibilities due to, e.g., asepsis requirements. In such cases, having the ability to see or fade out certain elements by simply focusing on the region of interest might be a valuable technique. On the other hand, the focus-dependent approach is mentally more demanding compared to alternatives such as semi-transparency or hiding/showing an object via user input. Although the fatigue results of the NASA-TLX questionnaire are quite reasonable, we expect that a longer, active use of the described switching between highlighted target and background might lead to significant exhaustion. As that phenomenon is less related to the highlighting aspect of Deadeye per se, we would rather consider a more fine-grained exploration as possible follow-up research.  

Ultimately, the question remains whether one should pick Deadeye or one of the more common preattentive techniques to draw and guide attention in VR visualizations. According to our qualitative evaluation, people prefer color over Deadeye in the case of greyscale scenarios. In our opinion, the most important advantage of color is its nonbinary nature, i.e., we can encode additional information into the highlighted target and differentiate between multiple targets by utilizing diverse colors. However, many visualizations are not greyscale, and applying our technique in such cases has clear advantages over other preattentive approaches: Deadeye does not alter any visual properties such as size or shape; performs well in complex, heterogeneous environments; and is straightforward to implement in stereoscopic setups. Compared to temporal approaches such as flickering, Deadeye does not suffer from typical VR-related issues such as aliasing due to constant movements of the HMD. To summarize, even though color is still superior in greyscale use cases, we emphasize the applicability and usefulness of Deadeye in colored scenarios and consider the technique as a viable addition to our visualization toolbox.


\section{Conclusion and Future Work}

This work explored the performance of Deadeye in virtual environments. Initially, Deadeye was introduced as a preattentive technique for drawing attention in 2D visualizations by removing the object of interest for one eye only. The major drawback for real-world usage was the requirement for extra stereoscopic equipment for highlighting, without taking any further advantage of such a setup. Hence, as a natural next stage of our research, we have now applied Deadeye in stereoscopic environments to study its interplay with 3D visualizations.

The paper made two contributions. First, the results from our preattentive tests with homogeneous and heterogeneous distractors confirmed that Deadeye behaves preattentively in VR, and, even more important, maintains its robustness in more complex setups with distractors varying in significant visual properties such as depth, color, and shape. Second, we demonstrated how Deadeye can be integrated into VR visualizations using the example of volume rendering. The outcomes of the exploratory study underpinned the highlighting capabilities of Deadeye and confirmed our ability to estimate depth of objects highlighted in this manner. Especially in colored visualizations, Deadeye was perceived as a valuable extension to our visualization toolkit, as the technique does not alter any visual properties, maintains robustness under common issues such as aliasing (``shimmering''), and comes with a straightforward implementation.

For our future work, we are particularly motivated by an observation during the presented volume rendering study: our participants noticed that the highlighted object could be faded in or out depending on focus, i.e., concentrating on the objects behind the target allows the user to suppress the target completely and see the background only, and vice versa. As possible follow-up research, we suggest investigating this multistable perception phenomenon in more detail to show its full potential for scientific visualizations beyond pure highlighting.

\acknowledgments{We are immensely grateful to Christine Pickett for her comments that greatly improved the manuscript. We would also like to show our gratitude to the anonymous reviewers of the original Deadeye paper and the IEEE VIS community for motivating us to explore Deadeye in virtual reality environments. This research was made possible in part by Award Number R01NR014852 from the NINR - National Institute of Nursing Research. }

\bibliographystyle{abbrv-doi}

\bibliography{deadeye}
\end{document}